\documentclass[aps,superscriptaddress,nofootinbib,balancelastpage]{revtex4}
\usepackage{epsfig,amsmath,amssymb,bm}
\def\bea{\begin{eqnarray}}
\def\eea{\end{eqnarray}}
\def\be{\begin{equation}}
\def\ee{\end{equation}}

\begin{document}
\title{Local Group dSph radio survey with ATCA (III): Constraints on Particle Dark Matter}

\author{Marco Regis}
\email{regis@to.infn.it}
\affiliation{Dipartimento di Fisica, Universit\`{a} di Torino, via P. Giuria 1, I--10125 Torino, Italy}
\affiliation{Istituto Nazionale di Fisica Nucleare, Sezione di Torino, via P. Giuria 1, I--10125 Torino, Italy}

\author{Sergio Colafrancesco}
\affiliation{School of Physics, University of the Witwatersrand, Johannesburg, South Africa}

\author{Stefano Profumo}
\affiliation{Department of Physics, University of California, 1156 High St., Santa Cruz, CA 95064, USA}
\affiliation{Santa Cruz Institute for Particle Physics, Santa Cruz, CA 95064, USA}

\author{W.J.G. de Blok}
\affiliation{Netherlands Institute for Radio Astronomy (ASTRON), Postbus 2, 7990 AA Dwingeloo, The Netherlands}
\affiliation{Astrophysics, Cosmology and Gravity Centre, Department of Astronomy, University of Cape Town, Private Bag X3, Rondebosch 7701, South Africa}
\affiliation{Kapteyn Astronomical Institute, University of Groningen, PO Box 800, 9700 AV, Groningen, The Netherlands}

\author{Marcella Massardi}
\affiliation{INAF - Istituto di Radioastronomia, Via Gobetti 101, I-40129, Bologna, Italy}

\author{Laura Richter}
\affiliation{SKA South Africa, 3rd Floor, The Park, Park Road, Pinelands, 7405, South Africa}

\date{Received \today; published -- 00, 0000}

\begin{abstract}
We performed a deep search for radio synchrotron emissions induced by weakly interacting massive particles (WIMPs) annihilation or decay in six dwarf spheroidal (dSph) galaxies of the Local Group.
Observations were conducted with the Australia Telescope Compact Array (ATCA) at 16 cm wavelength, with an rms sensitivity better than 0.05 mJy/beam in each field.
In this work, we first discuss the uncertainties associated with the modeling of the expected signal, such as the shape of the dark matter (DM) profile and the dSph magnetic properties.
We then investigate the possibility that point-sources detected in the proximity of the dSph optical center might be due to the emission from a DM cuspy profile.
No evidence for an extended emission over a size of few arcmin (which is the DM halo size) has been detected. We present the associated bounds on the WIMP parameter space for different  annihilation/decay final states and for different astrophysical assumptions.
If the confinement of electrons and positrons in the dSph is such that the majority of their power is radiated within the dSph region, we obtain constraints on the WIMP annihilation rate which are well below the thermal value for masses up to few TeV.
On the other hand, for conservative assumptions on the dSph magnetic properties, the bounds can be dramatically relaxed.
We show however that, within the next 10 years and regardless of the astrophysical assumptions, it will be possible to progressively close in on the full parameter space of WIMPs by searching for radio signals in dSphs with SKA and its precursors.
\end{abstract}

\maketitle

\section{Introduction}
\label{sec:intro}
Dark Matter (DM) is one of the defining issues and long-standing mysteries of modern physics.
Nowadays, there is an overwhelming amount of data, coming from cosmological, large-scale-structure, and galactic-scale observations, which lead us to conjecture that
more than 80\% of the total matter content of the Universe is most probably in the form of a not-yet-identified non-baryonic component, DM.

DM has been inferred so far through its gravitational properties only. There are however several scenarios in which DM is expected
to couple to ordinary matter, with the currently most investigated scheme postulating weakly interacting massive particles (WIMPs). 
The WIMP scenario (see, e.g., Ref.~\cite{Kolb:1990vq} for a review of key WIMP properties) is particularly appealing since it naturally predicts a DM relic density matching the observed cosmological DM density, and 
numerous extensions to the Standard Model of particle physics feature viable WIMP DM candidates (the lightest neutralino in SUperSYmmetry being the prime example).
WIMPs have masses in the GeV-TeV regime and can annihilate in pairs into lighter particles with a rate of the order of the weak interaction rate (with the canonical value needed to match the observed DM relic density for a thermally produced candidate being $\langle \sigma_a v\rangle=3\cdot 10^{-26} {\rm cm}^3/{\rm s}$).

Indirect detection studies have mainly been focused on the search for a WIMP-induced component in the neutrino and local antimatter cosmic-ray fluxes, and for an excess in
the multi-wavelength galactic or extra-galactic emissions (see, e.g., \cite{Bertone:2010at}).
Electrons and positrons can be directly or indirectly injected by WIMP annihilations. In particular, and with the exception of WIMP models annihilating exclusively into neutrinos, a sizable final branching ratio of annihilation into $e^+-e^-$ is a general feature of WIMP models (see e.g. Fig. 4 in Ref.~\cite{Regis:2008ij}). 
Emitted in environment with background magnetic fields, high-energy electrons and positrons give rise to a synchrotron radiation extending typically from the radio to the infrared bands.

Dwarf spheroidal (dSph) galaxies have been recognized as optimal laboratories for indirect DM searches \cite{Colafrancesco:2006he}.
They are not only the closest (other than the Galaxy itself) and most DM dominated objects in the local Universe, but they are also the faintest and most metal-poor stellar systems known (see, e.g., Ref.~\cite{McConnachie:2012vd} for a recent review). 
The presence of gas is tightly constrained by limits on HI emission~\cite{Grcevich:2009gt}, confirming the lack of recent star formation. 
No evidence for diffuse thermal or non-thermal emissions at X-rays and gamma-rays has been reported, implying very low quantities of
either hot ionized and relativistic plasmas.
The expected large DM signal, on one side, and low astrophysical background, on the other, make local dSphs an ideal target where to search for a WIMP signature.

Various attempts have been pursued in this context to detect WIMP-induced prompt emission of gamma-rays or a radiative emission (inverse Compton (IC) scattering in the X- and gamma-ray bands and synchrotron radiation at radio frequencies) associated with WIMP-induced electrons and positrons.
No evidence of diffuse signal has been obtained so far at any relevant frequency and this only allowed to set upper limits on the DM annihilation/decay rate for a wide range of WIMP masses.

The greatest effort for DM searches in dSphs has been so far undertaken by means of observations with gamma-ray telescopes.
Recently, the Fermi-LAT Collaboration have reported on 4-year observations of 25 dSph satellite galaxies of the Milky-Way (MW) \cite{Ackermann:2013yva}.
None of the targets have been detected with presented gamma-ray flux upper limits extending between 500 MeV and 500 GeV.
Constraints on the DM annihilation cross section have been derived from a joint likelihood analysis of a subset of 15 dSphs. At 95\% C.L. and assuming an NFW dark matter distribution, the bounds constrain the ``thermal'' annihilation rate $\langle \sigma_a v\rangle=3\cdot 10^{-26} {\rm cm}^3/{\rm s}$ for masses $\lesssim 10$ GeV in the cases of $\tau^+-\tau^-$ and hadronic final states (while being an order of magnitude weaker for lighter leptonic final states).
This study updates results from a 2-year analysis involving 10 dSphs~\cite{Ackermann:2011,Abdo:2010}, with bounds compatible within about a factor of 2.

TeV searches towards the most promising dSphs have been performed in the past ten years with ground-based Cherenkov telescopes, such as
Whipple~\cite{Wood:2008hx}, MAGIC~\cite{Albert:2007xg,Aliu:2008ny,Aleksic:2011jx,Aleksic:2013xea}, VERITAS~ \cite{Acciari:2010ab,Aliu:2012ga}, and HESS~\cite{Aharonian:2007km, Aharonian:2008dm,Abramowski:2010aa}. This technique is most sensitive to heavy WIMPs (with mass above 100 GeV or so) and the derived exclusion limits are at the level of $\langle \sigma_a v\rangle\simeq 10^{-23}-10^{-24} {\rm cm}^3/{\rm s}$ for DM annihilating into quarks and $\tau^+-\tau^-$. Recent analyses making use of gamma-ray data to place constraints on the WIMP DM parameter space include \cite{Bringmann:2008kj,Essig:2009jx,Scott:2009jn,Perelstein:2010at,Essig:2010em,Charbonnier:2011ft,Walker:2011fs,GeringerSameth:2011iw,Cholis:2012am} (see also references therein).

The population of non-thermal electrons and positrons induced by DM annihilations can be tested by means of X-ray observations of IC scattering on the CMB. Future observations can have interesting prospects in this respect ~\cite{Calvez:2011,Jeltema:2011bd}. Archival data of the XMM-Newton telescope in the fields of Ursa Minor, Fornax, and Carina dSphs were used by Ref.~\cite{Jeltema:2008ax} to constrain the WIMP parameter space. Bounds are at the level of $\langle \sigma_a v\rangle=10^{-22}-10^{-23} {\rm cm}^3/{\rm s}$ for $M_\chi=10$ GeV, assuming a MW-like spatial diffusion and an NFW DM profile. 

Radio observations of dSph galaxies have been also very recently attempted through the Green Bank Telescope (GBT) \cite{Spekkens:2013ik,Natarajan:2013dsa}, aiming to test the prediction of \cite{Colafrancesco:2006he}.
The FoVs of Draco, Ursa Major II, Coma Berenices, and Willman I were mapped at 1.4 GHz with a resolution of 10 arcmin and a sensitivity of 7 mJy/beam (after discrete source subtraction).
No significant emission was detected from the dSphs, and Ursa Major II and Willman 1 were found to be the cleanest cases to be exploited for setting bounds on the WIMP parameter space.
For a scenario where the DM annihilates into $b-\bar b$, the DM spatial distribution follows an NFW profile, the magnetic field strength is $B=1\,\mu$G, and the cosmic-ray particle diffusion is described as in the case of the Galaxy, Ref.~\cite{Spekkens:2013ik} found $\langle \sigma_a v\rangle<2.5\cdot 10^{-24} {\rm cm}^3/{\rm s}$ for $M_\chi=100$ GeV at 95\% C.L. (for a more extended discussion, see Ref.~\cite{Natarajan:2013dsa}).

In this paper (Paper III), we present the results of the search for WIMP radio signatures in six dSphs (Carina, Fornax, Sculptor, BootesII, Hercules and Segue2) making use of observations performed with the Australia Telescope Compact Array (ATCA) operating in the 1.1-3.1 GHz band.
In Section~\ref{sec:obs}, the observing setup and data reduction process are summarized (see Paper I~\cite{Paper1} and Paper II~\cite{Paper2} for further details). 
Essentially, the dSph targets were observed with both a compact core and long baselines, with sensitivity below 0.1 mJy (0.05 mJy) for a synthesized beam of 1 arcmin (6 arcsec).
We describe the models adopted for the expected WIMP signal (involving both microscopic properties and the DM spatial profile) and for the dSph interstellar medium in Section~\ref{sec:mod}.
The possibility of a detection of point-like emissions induced by WIMPs is investigated in Section~\ref{sec:detect}.
After subtracting point-sources, no firm evidence for an extended emission over the dSph size (few arcmin) was obtained. In Section~\ref{sec:bound}, we present the associated bounds on the WIMP parameter space, for different final states of annihilation/decay and for different astrophysical assumptions.
In Section~\ref{sec:compar}, these constraints are compared with the bounds on WIMPs in dSphs obtained by other experiments summarized above.
An outlook of the possibilities offered by the next-generation radio telescopes, which can significantly improve the sensitivity to WIMP signatures, is described in Section~\ref{sec:prospects}.
Section~\ref{sec:concl} summarizes our conclusions.

\section{Observations and data reduction}
\label{sec:obs}

The observations employed in this paper were performed during July/August 2011 with the six 22-m diameter ATCA antennae operating in the frequency range 1.1-3.1 GHz.
The targets included three classical dSphs (CDS), Carina, Fornax, and Sculptor, and three ultra-faint dSphs (UDS), BootesII, Hercules and Segue2, for a total of 123 hours of observing time.
The array configuration was formed by a core of five antennae with maximum baseline of about 200~m, and a sixth antenna located at about 4.5 km from the core. 
Details about the observing setup are provided in Paper I.

The data were reduced using the {\it Miriad} data reduction package~\cite{Sault:95}, following a standard calibration scheme.
Since we performed interferometric observations, data are collected in terms of spatial coherence of radiation measured at two points, or, in other words, in terms of correlated visibilities.
They consist of complex quantities made by the amplitude and phase of the cross-product of measured voltages from two antennae. 
The visibility function is a Fourier transform of the sky brightness distribution which can be inverted to produce an image of the sky. 
The visibility plane is also sometime called $UV$-plane with $U$ and $V$ being its coordinates.
Because of the duality between image and visibility planes, short distances in an image correspond to large spacings in the $UV$-plane, and viceversa.
At the central wavelength of observation ($\lambda\simeq16$ cm), such relation can be seen as $\theta= \lambda/b\simeq 5.5 (b/100\rm m)^{-1}$ arcmin, with $\theta$ being the angle in the sky-image and $b$ being the baseline length.

In the imaging process, we produced two types of maps.\footnote{Maps and source catalogue presented in this project can be retrieved at http://personalpages.to.infn.it/$\sim$regis/c2499.html.}
First, data were imaged in high-resolution maps (where the short baselines of the core are down-weighted) which can probe scales from few arcsec to few tens of arcsec, and have an rms noise of 30-40 $\mu$Jy.
In the second set of maps, we apply a Gaussian taper of 15 arcseconds (which down-weighs long baselines involving the sixth antennae) to the data before Fourier inversion.
The synthesized beam of these maps becomes about 1 arcmin, and the largest scale which can be well imaged is around 15 arcmin. Because of confusion limitation, the rms noise raises up to 0.1-0.15 mJy.
The main properties of the maps are summarized in Table 1 of Paper II. Details about the data reduction can be found in Paper I.

The tapered maps have a better sensitivity for testing the presence of a truly diffuse emission on scales above few arcmin, because of their larger beam.
For these maps, point sources (i.e., discrete sources in the high resolution maps) are a ``background'' that can be subtracted.
We perform this subtraction before Fourier inversion (namely, in the visibility plane) with the CLEAN component of the corresponding high-resolution map used as the input source model. This procedure is described in Sec.~3.1.1 of Paper II, and allows to bring the rms of the tapered map down by a factor of few (see Fig.~5 of Paper II). We will refer to these resulting images as MAP1.
However, some residuals are left and they are mostly due to a non-perfect reconstruction of the synthesized beam shape.
In order to remove them, a further subtraction can be performed in the final image of the tapered maps by removing point-like structures (which can be, in fact, remnant of a non-perfect subtraction in the visibilities), as described Sec.~3.1.2 of Paper II. The latter procedure (whose output will be called MAP2) might however be somehow risky since the beam of the map (and thus the size of point-sources) is about 1 arcmin, while the emission we are looking for is just a factor of few larger in size. In order to be conservative, when deriving bounds, we will mostly refer to maps with sources subtracted in the visibility plane only. 

To understand the impact of different imaging/subtraction methods on the intensity bounds, see also Fig.~6 in Paper II.

\begin{figure}[t]
\hspace{-30mm}
 \begin{minipage}[htb]{7.cm}
   \centering
   \includegraphics[width=0.55\textwidth,angle=-90]{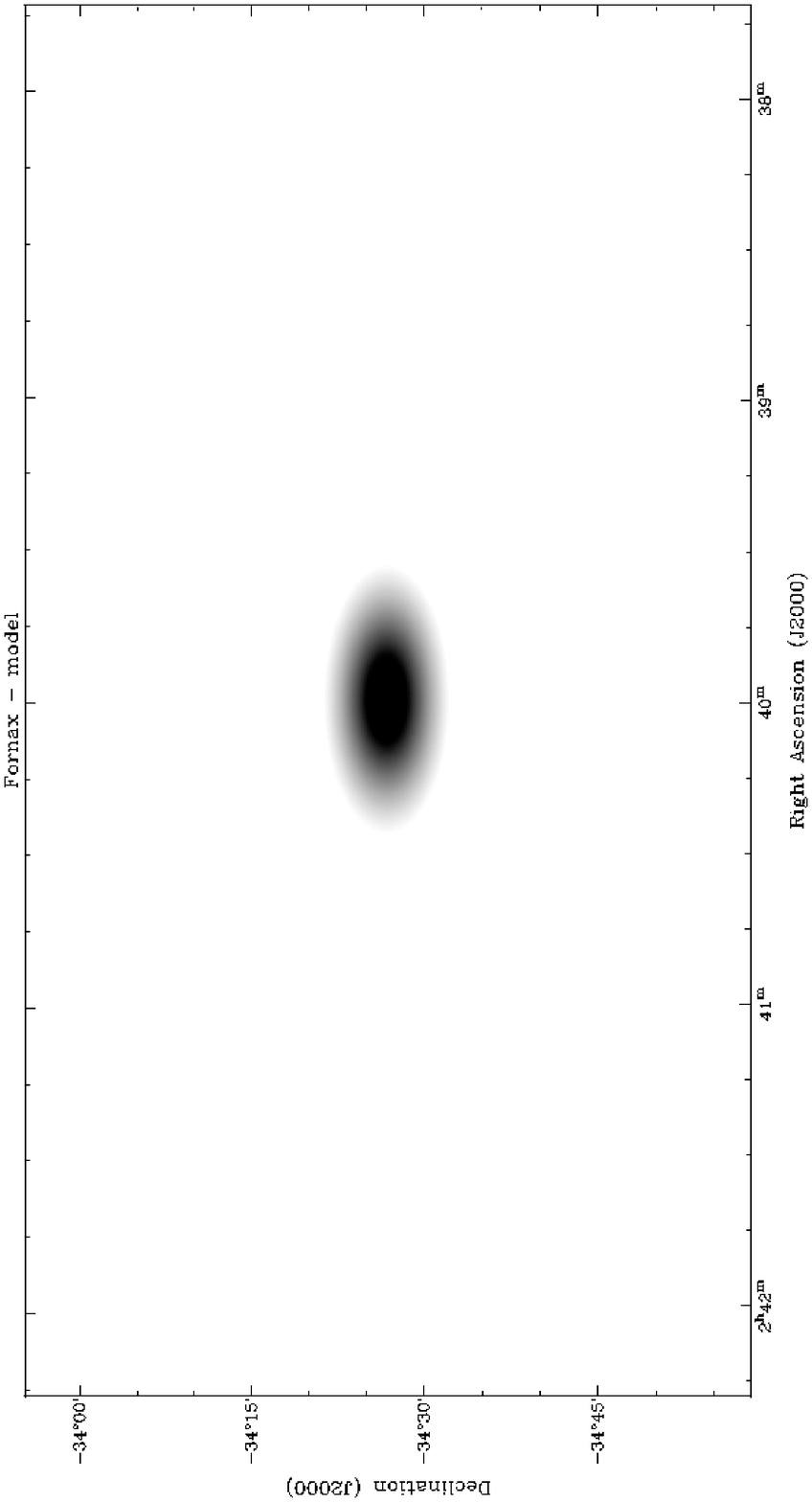}
 \end{minipage}
\hspace{0mm}
 \begin{minipage}[htb]{7cm}
   \centering
   \includegraphics[width=0.9\textwidth,angle=-90]{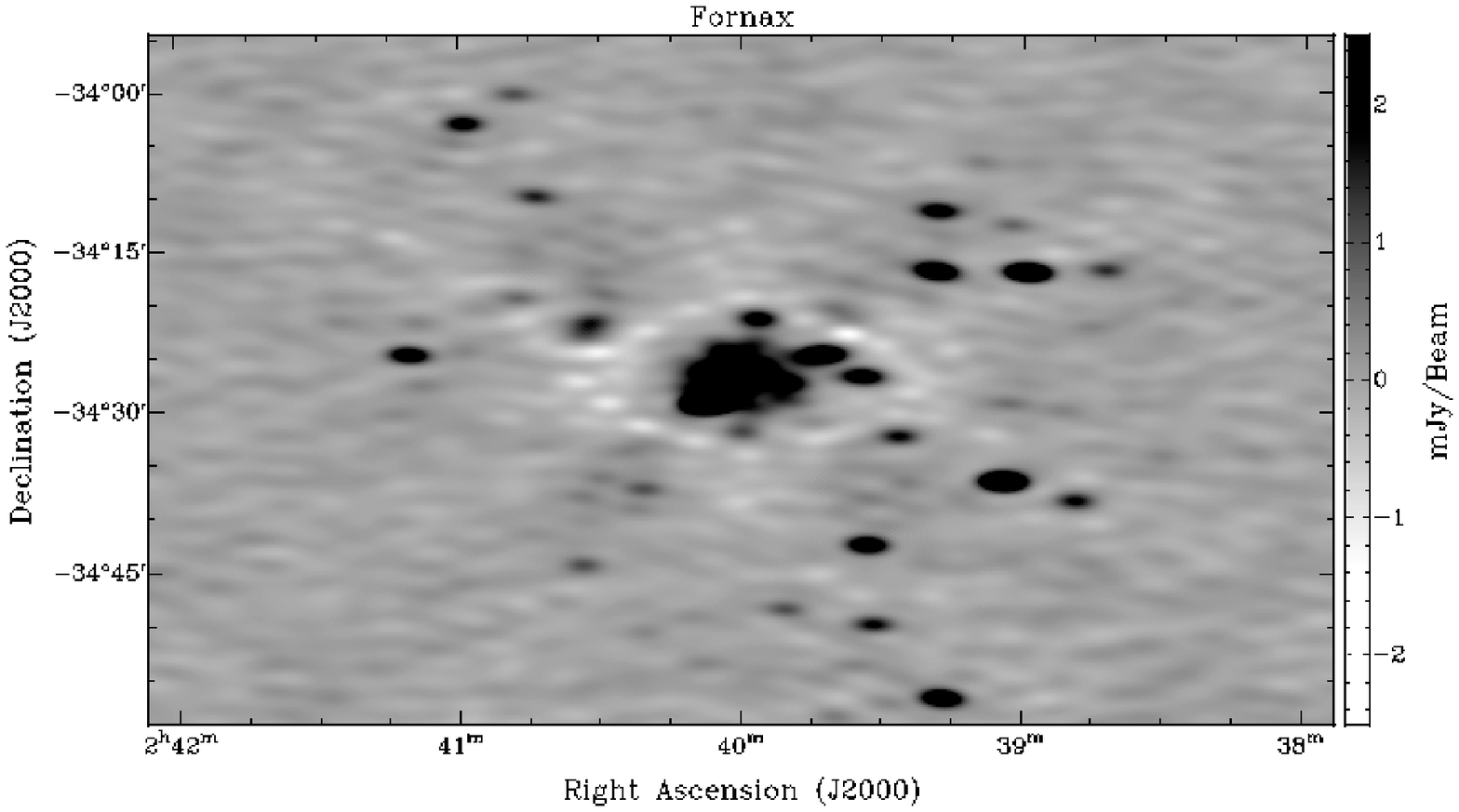}
 \end{minipage}
    \caption{Left: Expected signal for a WIMP with mass $M_{\chi}=100$ GeV and annihilation rate $\langle \sigma_a v\rangle=10^{-25} {\rm cm}^3/{\rm s}$ in the Fornax dSph. In this example, we consider a Burkert profile, $B_0=1\,\mu$G, and neglect spatial diffusion. Right: Map obtained adding the theoretical emission of the left panel to the visibilities of observational data, and then performing calibration and imaging as for the original map.}
\label{fig:fake}
 \end{figure}

In Fig.~\ref{fig:fake}, we show an example of the expected theoretical signal (left panel) and how it would have been reconstructed in our maps (right panel).
For illustrative purposes, we focus on the example of the Fornax dSph.
The simulated map is obtained by Fourier transforming the theoretical emission of the left panel, and adding the outcome to the actual data in the visibility plane.
This procedure has been performed by means of the task UVMODEL in {\it Miriad}.
After the simulated UV-data are derived, we reduced and imaged them following the procedure adopted to build the original maps (described in Section III of Paper I).
Fig.~\ref{fig:fake}b can be compared with the original map in Fig. 1 of Paper II.
In Section IIIb of Paper II, we demonstrated that the kind of signals we are going to discuss in the rest of the paper can be actually detected (within expected error bars) in our maps, by following the analysis pipeline we adopted. 

\section{Models}
\label{sec:mod}

\subsection{Dark Matter}
\label{sec:DM}

In the standard WIMP scenario, WIMP particles are stable but can annihilate in pairs injecting given species of particles (here we are interested in $e^+/e^-$) .
The DM WIMP source scales with the number density of WIMP pairs locally in space, i.e. assuming a smooth (without substructures), spherically symmetric, and static dark matter distribution, with $\rho^2/2\,M_{\chi}^2$, where $\rho(r)$ is the halo mass density profile at the radius $r$, and $M_{\chi}$ is the mass of the DM particle.
The source term associated to the production of $e^+/e^-$ is given by:

\be
q^a_e(E,r)=\langle\sigma_a v\rangle\,\frac{\rho(r)^2}{2\,M_{\chi}^2} \times \frac{dN_e^a}{dE}(E) \;,
\label{eqQ}
\ee
where $\langle \sigma_a v\rangle$ is the velocity-averaged annihilation rate, and $dN_e^a/dE$ is the number of electrons/positrons emitted per annihilation in the energy interval $(E,E+dE)$, obtained by weighting spectra for single annihilation channels over the corresponding branching ratio.
\footnote{In the case of WIMP as a Dirac fermion, the overall factor becomes 1/4, while 1/2 is appropriate for the more common cases of WIMP as a boson or Majorana fermion.}

Another possibility is that DM particles have a long but finite lifetime, and electrons and positrons are injected in DM decays.
In this case the source function takes the form:
\be
q_e^d(r,E)=\Gamma_d\,\frac{\rho(r)}{M_{\chi}} \times \frac{dN_e^d}{dE}(E) \;,
\label{eqQ2}
\ee
where $\Gamma_d$ is the decay rate and $dN_e^d/dE(E)$ is the number of electrons/positrons emitted per decay in $(E,E+dE)$.

The DM density profile $\rho$ in dSph galaxies is typically reconstructed from star kinematic data, as mentioned below.
All the other quantities in Eqs.~\ref{eqQ} and \ref{eqQ2} are instead related to the particle physics properties of DM, and 
they are the main subject of investigation in this paper.

\begin{table*}
\centering
\begin{tabular}{|l|c|c|c|c|c|c|c|}
\hline
dSph & distance &\multicolumn{2}{|c|}{NFW}&\multicolumn{2}{|c|}{Burkert}&\multicolumn{2}{|c|}{Einasto}  \\
name &$D$ [kpc]& $r_0$ [kpc] & $\rho_0$ [$M_\odot \,{\rm pc}^{-3}$]& $r_0$ [kpc] & $\rho_0$ [$M_\odot \,{\rm pc}^{-3}$]& $r_0$ [kpc] & $\rho_0$ [$M_\odot \,{\rm pc}^{-3}$]\\
\hline
Carina & 105 & $0.21$ & $0.30$ & $0.063$ & $3.3$ & $0.30$ &$3.7\cdot 10^{-2}$ \\ 
Fornax & 147 & $0.47$ & $0.13$ & $0.19$ & $0.91$ & $0.58$ &$2.2\cdot 10^{-2}$ \\ 
Sculptor & 86 & $0.39$ & $0.17$ & $0.13$ & $1.4$ & $0.47$ &$2.8\cdot 10^{-2}$ \\ 
BootesII & 42 & $0.17$ & $0.42$ & $0.052$ & $4.0$ & $0.23$ &$5.3\cdot 10^{-2}$ \\ 
Hercules & 132 & $0.20$ & $0.36$ & $0.060$ & $3.4$ & $0.26$ &$4.9\cdot 10^{-2}$ \\ 
Segue2 & 35 & $0.16$ & $0.44$ & $0.048$ & $4.3$ & $0.22$ &$5.4\cdot 10^{-2}$\\ 
\hline
\end{tabular}
\caption{Properties of DM profile. $D$ is the dSph distance taken from \cite{McConnachie:2012vd}, while $\rho_0$ and $r_0$ are the halo normalization and scale radius, respectively, derived from results in \cite{Martinez:2013els}.}
\label{tab:DMprof}
\end{table*}

\subsubsection{DM spatial profile}
\label{sec:DMprof}
The common way of deriving the DM profile in dSph assumes that dSphs are spherically symmetric systems at dynamic equilibrium and consisting of two components, a totally pressure-supported stellar population and the DM.
Strong tidal interactions can cause a departure from this picture. However, the data available for our sample of targets (which include Carina, Fornax, and Sculptor) do not support this possibility, and point towards low ellipticities (of order of 10-20\%~\cite{Lokas:2011sy}).
Within such framework, a relatively simple Jeans equation can be used to describe the system. Observational data about the line-of-sight velocity dispersion of stars can be then exploited to extract the mass profile (see, e.g.~\cite{Lokas:2001gy}).
The comparison with data is often performed through a Bayesian approach (including a certain set of priors for the quantities involved in the computation), and taking a Gaussian likelihood for the line-of-sight velocity dispersion of stars binned with respect to the projected radius.
With this approach and assuming a certain functional form $g$ for the DM profile, one can derive the most probable ranges for the halo normalization $\rho_0$ and scale radius $r_0$ of $\rho_{DM}(r)=\rho_0\,g(r/r_0)$.
Due to the well-known degeneracy between mass profile and the velocity anisotropy profile, dSph kinematic data can hardly distinguish between cored and cuspy DM halos, which both fit well the observational data.
We will consider one case of cored DM profile (Burkert), and two cases of cuspy DM profiles (NFW and Einasto):
\be
\rho_{BUR}=\frac{\rho_0}{(1+r/r_0)\,(1+r^2/r_0^2)}\;\;,\;\;\rho_{NFW}=\frac{\rho_0}{(r/r_0)\,(1+r/r_0)^2}\;\;,\;\; \rho_{EIN}=\rho_0\,\exp\left[-\frac{2}{\alpha}\left((\frac{r}{r_0})^\alpha-1\right)\right]\;.
\label{eq:prof}
\ee

The values of the halo parameters are taken from Ref.~\cite{Martinez:2013els}, by converting the the maximum circular velocity $v_{max}$ and radius corresponding to this velocity $r_{max}$ quoted in their Table 2 in terms of the halo normalization $\rho_0$ and scale radius $r_0$ used in the present work. The parameter $\alpha$ in the Einasto profile has been set to $\alpha=0.15$ from the best-fit value in \cite{Martinez:2013els}.

In the case of BootesII, a robust determination of dynamical properties is hindered by the fact that only few tens of stars have been observed in its FoV~\cite{Walsh:2007tm}, and the observed velocity dispersion is quite uncertain~\cite{Koch:2008wx}. Structural properties seem however to suggest a picture in between other two ultra-faint dSphs, Coma Berenices and WillmanI~\cite{Martin:2008wj}. For simplicity, and, in order to have an estimate of the order of magnitude for the expected signal, we compute $\rho_0$ and $r_0$ for BootesII, by averaging the values of $v_{max}$ and $r_{max}$ of Coma Berenices and WillmanI in \cite{Martinez:2013els} (which differ very little from each other).

\subsubsection{Particle physics properties}
\label{sec:DMpp}
We  derive constraints on particle DM models in the plane $M_{\chi}$-$\langle \sigma_a v\rangle$ and $M_{\chi}$-$\Gamma_d$ for few different choices of $dN_e/dE$.
If the two-body final state particles from WIMP annihilations (decays) are predominantly quarks or weak gauge bosons, the injection of $e^+/e^-$ is mainly associated to a chain of hadronization and decay processes, leading to the production of charged pions and their subsequent decays into muons and in turn into $e^+/e^-$.
When instead the process of annihilation (decay) dominantly produces leptons, then $e^+/e^-$ are mainly originated directly from decays and have harder spectra and larger branching ratios.
We will consider $dN_e/dE$ as provided by $b-\bar b$ and $W^+-W^-$ final states (as illustrative for the first scenario), and by $\tau^+-\tau^-$ and $\mu^+-\mu^-$ final states (to describe the second case).

\begin{figure}[t]
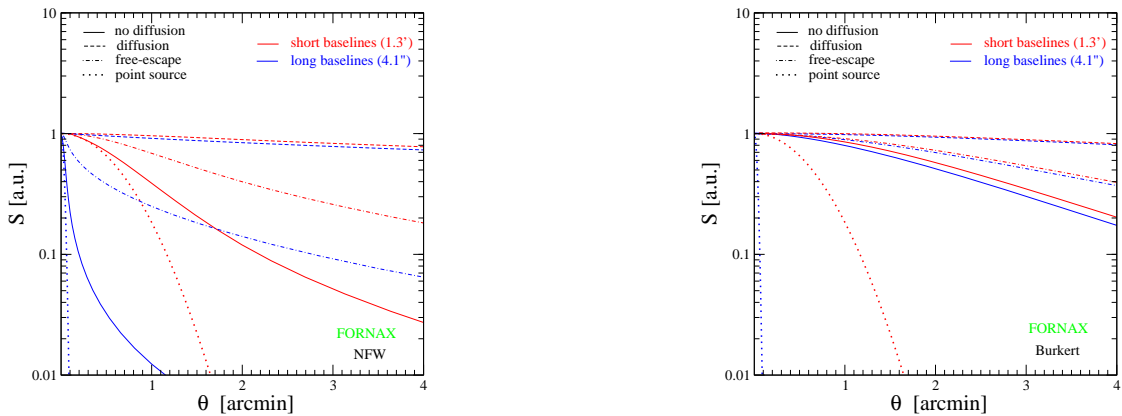

 \begin{minipage}[htb]{7.cm}
   \centering
   \includegraphics[width=0.8\textwidth]{DMsourcesize.eps}
 \end{minipage}
\hspace{20mm}
 \begin{minipage}[htb]{7cm}
   \centering
   \includegraphics[width=0.8\textwidth]{DMsourcesize_2.eps}
 \end{minipage}
    \caption{We show the expected angular profile for the DM source. The example of the Fornax dSph is taken as reference. The electrons and positron are modeled such that they emit at injection place (solid), diffuse with a constant $D$ given by $D=3\cdot 10^{28}\,(E/{GeV})^{1/3}{\rm cm^2/s}$ (dashed), or undergo free-escape (dashed-dotted), for an observational angular resolution corresponding to approximately (we take a circular beam for this plot) the tapered (1.3 arcmin, red) and un-tapered (4.1 arcsec, blue) Fornax maps. The case of a point-source (dotted) is shown for comparison. Arbitrary units are used and all the cases are normalized to have the same peak flux density. Left panel is for an NFW profile, while in the right panel a Burkert profile is considered.}
\label{fig:DMsize}
 \end{figure}

\subsection{dSph interstellar medium}
\label{sec:dSph}
The transport of high-energy electrons and positrons injected by DM in the dSph can be described as a diffusive process governed by the equation:
\be
 -\frac{1}{r^2}\frac{\partial}{\partial r}\left[r^2 D\frac{\partial f_e}{\partial r} \right] 
  +\frac{1}{p^2}\frac{\partial}{\partial p}(\dot p p^2 f_e)=
  s_e( r, p)
\label{eq:transp}
\ee
where we assumed spherical symmetry and stationarity. In Eq.~\ref{eq:transp}, $r$ is the radius, $p$ is the momentum, and $f_e(r,p)$ is the equilibrium $e^+-e^-$ distribution function, related to the number density in the energy interval $(E,E+dE)$ by: $n_e(r,E)dE=4\pi \,p^2f_e(r,p)dp$. Analogously, for the DM source function, we have $q_e(r,E)dE=4\pi \,p^2\,s_e(r,p)dp$. 
The first term on the left-hand side describes the spatial diffusion, with $D(r,p)$ being the diffusion coefficient. The second term accounts for the energy loss due to radiative processes; we include synchrotron and inverse Compton on CMB.
Further details on the modeling and on the numerical solution of Eq.~\ref{eq:transp} can be found in Paper II.

The magnetic properties of dSphs are poorly known and this strongly affects the predictions for the synchrotron signal (for an extended discussion, see Sec.~4.3 in paper2).
Here, in order to bracket associated uncertainties, we define three benchmark scenarios summarized in Table~\ref{tableB}. 
In the first, termed ``optimistic'' (OPT), we derive the magnetic field from the equipartition condition, namely we impose $U_B(r)=3/4\,U_e(r)$ with $U_B(r)=B(r)^2/(8\,\pi)$ and 
$U_e(r)=\int_{0.1\,{\rm GeV}}^{M_{\chi}}dE\,E\,n_e(E,r)$ (we conservatively consider the electrons and positrons injected by DM to be the only CR component in the dSph). Furthermore, we assume the $e^+-e^-$ to radiate all of their energy at the same place of injection (i.e., disregarding diffusion), and the DM to be distributed following an Einasto profile. 

The second benchmark scenario is an ``average'' (AVE) model. The magnetic field is derived from the density of star formation rate $\Sigma_{SFR}$ in the dSph and assuming a scaling $B\propto\Sigma_{SFR}^{0.3}$ which has been found to hold in the Local Group~\cite{Chyzy:2011sw}. The spatial profile of $B$ is described by $B=B_0\,e^{-r/r_*}$, with $r_*$ being the stellar half-light radius, and we consider the SFR rate averaged over the history of the dSph (both $B_0$ and $r_*$ are reported in Table 2 of Paper II). If such estimate leads to a magnetic field strength below $1\,\mu$G, we assume $B_0=1\,\mu$G (which can be seen as an estimate of the magnetization of the MW surroundings). For more details, see Paper II.
In the AVE model, the diffusion coefficient is taken with normalization and spectrum defined in order to have a Milky-Way like diffusion within the stellar region and then growing exponentially in the outskirt: $D=3\cdot 10^{28}\,(E/{GeV})^{0.3}\exp(r/r_*)\,{\rm cm^2/s}$. The DM profile follows an NFW distribution.

Finally, in the ``pessimistic'' (PES) scenario, we take a magnetic field model similar to the AVE one, but with normalization derived from late-time SFR (see column 5 in Table 2 of Paper II) and without lower limit at $B_0=1\,\mu$G. A cored Burkert profile is assumed for the DM spatial distribution. The diffusion coefficient is increased by a factor of 30: $D=10^{30}\,(E/{GeV})^{0.3}\exp(r/r_*)\,{\rm cm^2/s}$. 
This diffusion scheme provides bounds which are only a factor $\mathcal{O}(1)$ stronger than in the ``free-escape'' case (i.e., with particles going out of the dSph at the speed of light without random-walk), see Fig.~11 in Paper II.
The latter case is however irrealistic, since CRs cannot stream along a magnetic field much faster than the Alfv\`en speed because they generate magnetic irregularities which in turn scatter them (see, e.g., \cite{Cesarsky:1980pm}). Even in the case of pure CR self-confinement, with no other ionized medium in the dSph rather than the DM-induced component and the cosmological density of electrons, a flux larger by a factor of few than in the "free-escape" scenario is expected. Thus the depicted PES scenario can be actually taken as the extreme reference case.

For all the three benchmark scenario, we set the parameters $r_0$ and $\rho_0$ to their central observational value reported in Table~\ref{tab:DMprof}.

\begin{table*}
\centering
\begin{tabular}{|c|c|c|c|}
\hline
Name & magnetic field &diffusion scheme & DM profile  \\
\hline
OPT & $B_{eq}^{obs}$ & loss-at-injection & Einasto \\ 
AVE & max($B_{\overline{SFR}},\,1\,\mu$G) &  $D=3\cdot 10^{28}\,(E/{\rm GeV})^{0.3}\exp(r/r_*)\,{\rm cm^2/s}$ & NFW \\ 
PES &$B_{SFR_0}$ & $D= 10^{30}\,(E/{\rm GeV})^{0.3}\exp(r/r_*)\,{\rm cm^2/s}$ & Burkert \\ 
\hline
\end{tabular}
\caption{Benchmark astrophysical scenarios. Columns shows the name of the model, the magnetic field (see Table 2 of Paper II), the diffusion scheme, and the DM profile.}
\label{tableB}
\end{table*}

\section{Detection}
\label{sec:detect}

If the DM spatial distribution follows a cuspy profile (like, e.g., an Einasto or NFW profile), as possibly favoured by N-body numerical simulations, with an eventual core of size much smaller than few pc (i.e., beyond our best angular resolution), and the $e^+-e^-$ emit synchrotron radiation at the same place of injection (i.e., without undergoing spatial diffusion), the WIMP source can appear as a point-like source located at the center of the dSph.
If, on the other hand, the DM profile is cored (like, e.g., for a Burkert profile) or the diffusion is relevant (i.e., the confinement time is sufficiently long to reshape a cuspy $e^+-e^-$ distribution into a shallower one), the dSph WIMP source will be extended over the size of few arcmin (which is the typical size of a possibly magnetized region in dSph).
The first scenario can be better tested in the high-resolution maps mentioned in Section~\ref{sec:obs}, such to use our best synthesized beam which is, on average, of about 6 arcsec.
In the second scenario, it is instead more promising to integrate the flux on arcmin scales, and therefore to make use of the tapered maps such that the synthesized beam becomes of about one arcmin (while long-baseline data are used to remove point-sources).

We illustrate these pictures in Fig.~\ref{fig:DMsize}, taking the example of the Fornax dSph.
In the case of no diffusion, the NFW profile (and similarly for the Einasto profile) can be resolved only for bright emissions, namely, with a peak well above the detection threshold, otherwise it appears as a point-like source (as can be understood by comparing dotted and solid lines in Fig.~\ref{fig:DMsize}a).
When diffusion is important, on the other hand, the emission profile becomes significantly reshaped and flattened (dashed lines).
The case of $e^+-e^-$ free-escape (dash-dotted lines) stands, for what concerns the spatial shape, somewhat in between the two above scenarios.
Note that all the cases are normalized to have the same peak flux density (while in physical units the free-escape scenario would clearly be the faintest case).
In Fig.~\ref{fig:DMsize}, we show the three different scenarios for both the angular resolutions of short and long baselines (i.e., tapered and un-tapered maps).

In the case of the Burkert profile, the source is instead extended in all the diffusion schemes and observing setups.
Note from the plots that, as the source becomes more diffused, and in particular for sizes much larger than both beams, the signals of the two different setups tend to coincide.
The other dSphs show similar properties (which in part is coincidentally due to the fact that closest objects (UDS) are also the smallest ones, so angular profiles do not change much).

In Fig.~\ref{fig:DMsize}, we show the signal in the long-baseline case up to few arcmin. This is done for illustrative purposes only, since it would be the case for observations having a synthesized beam of few arcsec, but including shorter baselines to be able to image large structures. The high-resolution map considered here is instead very weakly sensitive to source sizes above 30 arcsec. 

The plots confirm that physical cases such that the DM source size is $\ll$ arcmin (i.e., within the scales probed by the high-resolution maps and smaller than short-baseline resolution) can be better investigated using long-baselines maps. Indeed, they have a smaller synthesized beam with respect to the short-baseline case and are not confusion limited, thus having higher sensitivity.
In the next Section, we will discuss the point-source detection near the dSph centers and their possible association to emissions from cuspy DM profiles.
We will also consider constraints on DM point-like emission when discussing the OPT benchmark model.

On the contrary, the tapered map, due to its larger beam, is the map of choice for investigating DM scenarios inducing a signal on scales above the arcmin, as in the AVE and PES models.

Details about the statistical techniques for both detections and bounds (which are presented in the next Sections) can be found in Section 5.1 of Paper II.

\begin{figure}[t]
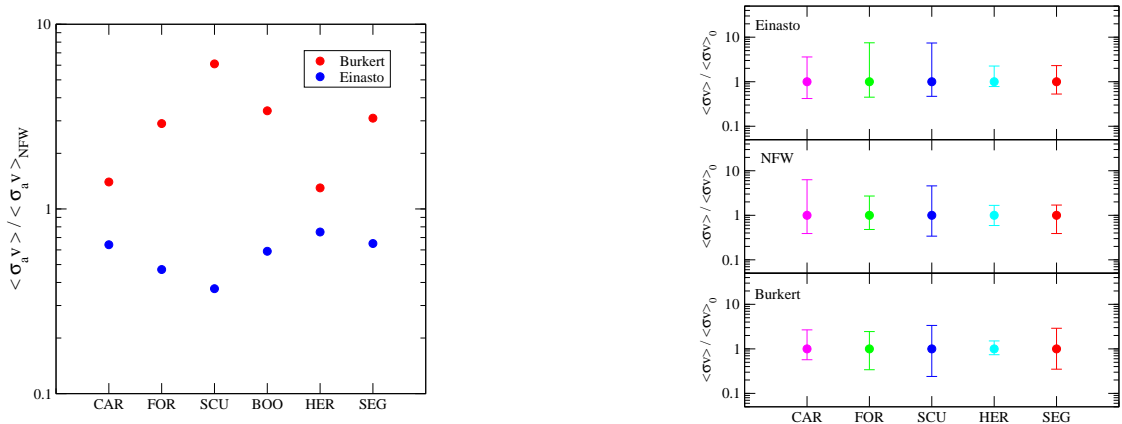

 \begin{minipage}[htb]{7.cm}
   \centering
   \includegraphics[width=0.8\textwidth]{DM_typeprof.eps}
 \end{minipage}
\hspace{20mm}
 \begin{minipage}[htb]{7cm}
   \centering
   \includegraphics[width=0.8\textwidth]{DM_parprof.eps}
 \end{minipage}
    \caption{Left: Ratio between the 95\% C.L. bound on $\langle \sigma_av\rangle$ obtained with a NFW versus Burkert and Einasto DM profiles.
  Right: Variations of the 95\% C.L. bound on $\langle \sigma_av\rangle$ by varying $r_0$ and $\rho_0$ over their observationally allowed range. In both figures, we considered a WIMP with $M_{\chi}=100$ GeV, annihilating into $b-\bar b$, and neglected spatial diffusion. See text for further details.}
\label{fig:boundr0rho0}
 \end{figure} 

\subsection{Point-sources near the dSph center and their possible association to a cuspy DM profile}
\label{sec:PS}

In order to explore the possibility of detecting an emission from the center of the DM profile, we consider both that the uncertainty in the centroid position of the dSph stellar distribution is about 1 arcmin and that the center of the DM distribution might be slightly offset from it. Thus we analyze all the sources within 2 arcmin from the dSph center. In principle, this includes also possible big DM clumps located in the central region of the DM profile.
Another possibility recently investigated in Ref.~\cite{Gonzalez-Morales:2014eaa} is related to the possible presence of an intermediate massive black holes (IMBH) hosted at or around the center of the dSphs. In this scenario, a signature stemming from a spiky DM density (due to adiabatic contraction in the IMBH gravitational potential) or from the accretion of the IMBH itself is expected.

In Table~\ref{tab:sources} we report the list of excesses at a statistical evidence of more than 3-$\sigma$. 
We searched for sources of different sizes (from point-like to the largest testable size in the high resolution map, i.e., few tens of arcsec) centered within the inner 2 arcmin of the maps, and then we fitted them with 2D elliptical Gaussians.
More precisely, we run the task SFIND of the {\it Miriad} package, setting the parameter $\alpha$ for the maximal false detection rate to 100. This roughly corresponds to a 3-$\sigma$ threshold for detection, and is a weaker criterium with respect to the one adopted in building the catalogue, which has been $\alpha=1$ (namely, around 5-$\sigma$; for more details about source extraction, see Paper I). 
Indeed, except for the two sources in the Carina FoV and for the brightest source in the Segue2 Fov, the other cases reported in Table~\ref{tab:sources} do not appear in the catalogue presented in Paper I.

None of the sources listed in Table~\ref{tab:sources} is located at the exact center of the dSph, and distances are well above the positional accuracy of our observing setup (which is at arcsec level). 
In the case of the BootesII dSph, we include a source at a distance slightly larger than $2'$, taking into account that the stellar population, and in turn its center, is currently poorly known.  

All the sources in Table~\ref{tab:sources} are compatible (within errors) with being point-like.
Therefore, the only DM scenario which is able to fit has to foresee a sufficiently cuspy spatial profile and radiation of $e^+-e^-$ at injection position.
We fit the reported excess taking an Einasto profile (since the sources are unresolved we cannot distinguish among different cuspy profiles) with parameters as in Table~\ref{tab:DMprof}. We assume a magnetic field strength as in the $B_{\overline{SFR}}$ scenario, and, in order to fix the ideas, choose $M_\chi=100$ GeV and $b-\bar b$ final state. 
Since the sources are point-like, the fit reduces to match one number, the level of the measured flux, which sets $\langle \sigma_a v\rangle$. The corresponding best-fit value is reported in the last column of Table~\ref{tab:sources}.

Intriguingly, only the sources in Carina and Hercules FoVs would require pretty large annihilation rates, while all the other values are well compatible with constraints from other DM indirect searches.
On the other hand, the derived distances from the dSph optical centers are above 1 arcmin; moreover, in order to have a source below few arcsec of size, a very strong confinement mechanism must be at work, to keep $e^+-e^-$ in a region smaller than 10 pc.

We perform a search on archival catalogues  (to this aim we use the web interface provided at www.asdc.asi.it, as well as queries to the NED) in order to identify sources that already have an association with a known astrophysical object type.
For these cases the DM interpretation can be consequently excluded.
We focus on typical radio-counterpart (i.e., IR, X-rays, gamma-rays), while obviously in the optical we would have found the old dSph stars, which however unlikely emit at radio frequencies.
Only the first source in the Table \ref{tab:sources} can be confidently associated to a known source (most probably, a star). In this case, a source with IR, UV and X-rays emission has been detected by WISE, GALEX and XMM at $\approx$ 13 arcsec from the position quoted in Table~\ref{tab:sources}. The source has been also detected in the USNO optical survey. The existing photometric data might suggest a possible correlation with an IMBH and we will explore this interesting possibility in more details elsewhere. 
In all the other cases, possible counterparts are at significantly larger distances, namely, they are not compatible after taking into account the positional accuracy of our survey and of the other analyzed surveys.

The DM interpretation, while not being a straightforward possibility (as discussed above), in most of the cases in Table~\ref{tab:sources} cannot be ruled out based on the available data.

\begin{table*}
\centering
\caption{Properties of point-sources located within 2 arcmin from the dSph optical centers. The values of $\langle \sigma_a v\rangle$ have been determined matching the measured flux density with the emission from a WIMP model with $M_\chi=100$ GeV, $b-\bar b$ final state, Einasto DM profile, $B_{\overline{SFR}}$ and no diffusion, see text.}
\label{tab:sources}
\begin{tabular}{|c|rrr|rrr|c|c|c|}
\hline
dSph &\multicolumn{6}{|c|}{J2000}& Distance  &  Flux density  & $\langle \sigma_a v\rangle$   \\
FoV &\multicolumn{3}{c}{RA} & \multicolumn{3}{c|}{DEC} & arcmin  & mJy  & $ 10^{-26} {\rm cm}^3/{\rm s}$\\
\hline
CAR & 06 & 41 & 33.5 & -50 & 58 & 11.7& 0.6 & $0.28 \pm 0.05$ & 11.9\\ 
CAR & 06 & 41 & 27.6 & -50 & 59 & 09.5& 1.9 & $0.30 \pm 0.05$ & 12.7\\ 
FOR & 02 & 40 & 00.3 & -34 & 25 & 07.6& 1.8 & $0.16 \pm 0.04$ & 1.1\\ 
SCU & 01 & 00 & 15.0 & -33 & 44 & 00.3& 1.9 & $0.28 \pm 0.06$ & 1.8\\ 
BOO & 13 & 58 & 04.2 &  12 & 52 & 53.6& 2.2 & $0.17 \pm 0.05$ & 0.51\\ 
HER & 16 & 31 & 00.2 &  12 & 46 & 48.1& 0.8 & $0.11 \pm 0.04$ & 11.7\\ 
SEG & 02 & 19 & 18.7 &  20 & 09 & 13.1& 1.4 & $0.09 \pm 0.03$ & 0.27\\ 
SEG & 02 & 19 & 18.0 &  20 & 11 & 39.1& 1.2 & $0.22 \pm 0.03$ & 0.66\\ 
\hline

\end{tabular}
\end{table*}

N-body numerical simulations suggest the DM distribution in structures to be clumpy.
Therefore, one could explore the possibility to detect single clumps in our maps.
In the source catalogue reported in Paper I, we have 1392 detected sources.
Even filtering the catalogue by discarding identified sources (i.e., with an association to a known astrophysical source at other wavelength) and sources lying outside the virial radius, we are still left with a large number of sources and it would be rather speculative to analyze all of them in terms of DM emission from subhalos. 

Moreover, although it is in principle not excluded to have small clumps appearing as point-sources, it is very likely that spatial diffusion would make their size much broader.
In this case, they will therefore rather contribute to the global extended emission of the dSph.
Assuming a certain subhalo distribution, one can in principle compute the additional component of the dSph diffuse emission given by DM annihilations in substructures.
On the other hand, the mass function of subhalos in dwarf galaxies is still very uncertain. N-body simulations do not currently reach the resolution to study the subhalo distribution for main halo mass $\lesssim10^8\,M_\odot$, as would be required in the case of dSphs. 
We conservatively decide not to include a substructure boost to the total diffuse emission of annihilating DM.
In summary, we neglect any possible contribution from subhalos.

\subsection{Diffuse emission}
\label{sec:diff}

In Paper II, we tested different spatial profiles for the diffuse emission in dSphs and, from the analysis of single dSphs, found no significant evidence.
Here we re-perform the test in the AVE and PES scenarios, by setting the halo parameters $r_0$ and $\rho_0$ to the value in Table~\ref{tab:DMprof}, and fitting all the dSph simultaneously with two free parameters ($M_\chi$ and $\sigma_a v$), since DM microscopic properties are considered to be universal.

Similarly to the results in Paper II, for the maps with point-sources subtracted in the visibility plane only (MAP1), we obtain a significant statistical evidence, at the level of 12-$\sigma$ for both the AVE and PES scenarios. Since our measurement concerns only one frequency bin and the injection spectrum does not have a great impact on the source morphology (as we will better discuss below), the fit is not able to break the degeneracy between mass and annihilation rate (which both enters in the normalization of the signal). Very mild preference for large masses is found with best-fit around $M_{\chi}=5$ TeV and $\langle \sigma_a v\rangle_{AVE}=4\cdot 10^{-23} {\rm cm}^3/{\rm s}$ ($\langle \sigma_a v\rangle_{PES}=1.6\cdot 10^{-21} {\rm cm}^3/{\rm s}$) for annihilations into $b-\bar b$.
However, as discussed in Paper II, the model is actually fitting a residual emission from sources (due to a non-perfect subtraction), rather than a truly diffuse emission.

After removing the remaining sources in the image plane of maps having already undergone source subtraction in the Fourier space (MAP2), the signal case is no longer preferred with respect to the null hypothesis.
Therefore no compelling evidence has been found. 
In Section~\ref{sec:bound}, we proceed to compute constraints on $\langle \sigma_a v\rangle$ and $M_{\chi}$.

To further explore the possibility of a detection, one can perform a stacking of the 6 dSph maps.
The stacking is not totally well-defined since we expect a different size for each dSph, and there is not robust scaling factor that can be adopted. However, the prediction of an excess in the inner few arcmin is common to all the cases.
We simply stacked together the 6 dSphs without any rescaling and fit the resulting map with a phenomenological form for the emissivity following either an NFW or a Burkert profile (with free $r_0$ and overall normalization of the signal).
The operation is performed for the two types of source-subtracted maps (i.e., MAP1 and MAP2). The results are similar as in the case of the combined fit discussed above, namely, a significant detection for the case with sources subtracted in the visibility plane only (due to remnants of point-sources) which disappears when considering maps with sources subtracted also in the image plane.

\section{Constraints on WIMP parameter space}
\label{sec:bound}

Before deriving constraints on the WIMP parameter space, we discuss the degree of uncertainty associated to the DM profile.
In Fig.~\ref{fig:boundr0rho0}a, the impact of the profile shape on the bounds is shown.
We report the ratios of the annihilation cross-sections for the Einasto and Burkert profiles over the NFW case. They are computed taking $\rho_0$ and $r_0$ from Table~\ref{tab:DMprof}.
The absence of a central cusp in the Burkert case makes constraints weaker, while the Einasto profile provides the most stringent constraints.
For this plot, we consider the no-diffusion case and the tapered map with sources subtracted in the visibilities.
The overall uncertainty due to the profile shape is between a factor of 2 to 15, depending on the dSph.
Note that this conclusion depends on the diffusion scenario adopted and on the type of map considered. For example, the presence of sources, especially near the center, and a noise which varies from the center to the outer part of the map, can have a significant impact on the ratio of the bounds. This is because a cuspy scenario is obviously more constrained by the inner data, while a more extended emission involves a comparison with the full map. 
However, for all the maps and diffusion scenarios considered here, the overall uncertainty stays roughly within one order of magnitude.

Note that in the case of Hercules (and to a lesser extent of Carina) the variation is quite limited. This is because the size of the dSph is relatively small, and the stellar velocity dispersion is relatively well-known. The latter allows a more precise determination of the DM profile with respect to the other UDS (BootesII and Segue2).
Fornax and Sculptor, due to their larger size, have instead a larger area involved in the fit which makes the dependence on the shape stronger than in the Hercules case (which, for the diffusion scheme and type of maps adopted in this plot, nearly resembles a point-source).

In Fig.~\ref{fig:boundr0rho0}b, we show the uncertainty on the cross-section bounds arising from the uncertainty on $\rho_0$ and $r_0$.
In order to derive the allowed range, we consider the 1-$\sigma$ band for $v_{max}$ and $r_{max}$ reported in Table 2 of \cite{Martinez:2013els} and assume they are fully correlated (we need to make an approximation since the covariance is not provided). 
In other words, we took the 1-$\sigma$ upper values of both $v_{max}$ and $r_{max}$, translate them into $\rho_0$ and $r_0$ and compute the upper value of $\langle \sigma_a v\rangle$ (shown in Fig.~\ref{fig:boundr0rho0}b), and similarly for the lower value, taking the 1-$\sigma$ lower values of $v_{max}$ and $r_{max}$.
To check our approximation, we computed the so-called J-factor, which is given by $\int_{\Delta \Omega}d\Omega \ \int_{los}dl\ \rho^2({\bf r})$. Taking $\Delta\Omega=2.4\cdot10^{-4}$ sr, our findings matches with the results reported in Table I of \cite{Ackermann:2013yva}, which were computed with a full implementation of the method of \cite{Martinez:2013els}.

Fig.~\ref{fig:boundr0rho0}b shows that the uncertainty related to the observationally allowed range of $\rho_0$ and $r_0$ is within about one order of magnitude. 
Again, this factor varies depending on the diffusion scenario adopted, but we checked that such variation turn always out to be relatively small.

The results of Fig.~\ref{fig:boundr0rho0} thus shows that the uncertainty related to the DM profile has a smaller impact on the bounds than the description of dSph magnetic properties.
On the other hand, we should bear in mind that the analysis of \cite{Martinez:2013els}, similarly to other works in the field, includes a number of priors, which somehow restrict the allowed ranges for the parameters, while more conservative and broader ranges could be in principle considered.

In order to bracket all the uncertainties associated to the required astrophysical inputs, we set three benchmark scenarios, OPT, AVE, and PES, which are listed in Table~\ref{tableB} and discussed in Section~\ref{sec:dSph}.
We compute bounds on the WIMP microscopic properties in the $M_\chi-\sigma_a v$ plane for different final states of annihilations/decays.
The likelihood is computed as described in in Section 5.1 of Paper II and focusing on the central dSph region, taking a radius of 30 arcmin (20 arcmin) for the CDS (UDS).
The relevant region for the statistical test will be actually related to the source area, since the part of the likelihoods outside this region cancels out in the likelihood ratio (being identical between different models). The source region can be significantly smaller than 20-30 arcmin, and at most corresponding to the largest achievable scale of the maps, which is of the order of 15 arcmin (30 arcsec) when considering low (high) resolution maps, see PaperII.
On top of single target analyses, we also perform a combined analysis where the likelihood is given by the product of the likelihoods of the each dSph.

As already mentioned, the uncertainty in the centroid position of the dSphs considered here is typically estimated to be below arcmin level.
On top of that, the center of the spherical DM distribution might be slightly offset with respect to the stellar distribution.
However, since our sensitivity is rather homogeneous on scales comparable to the dSph size, we do not expect a significant variation of the bounds due to possible misalignment between the assumed center of the spherical distribution of our models and the real dSph DM center.

\begin{figure}[htb]
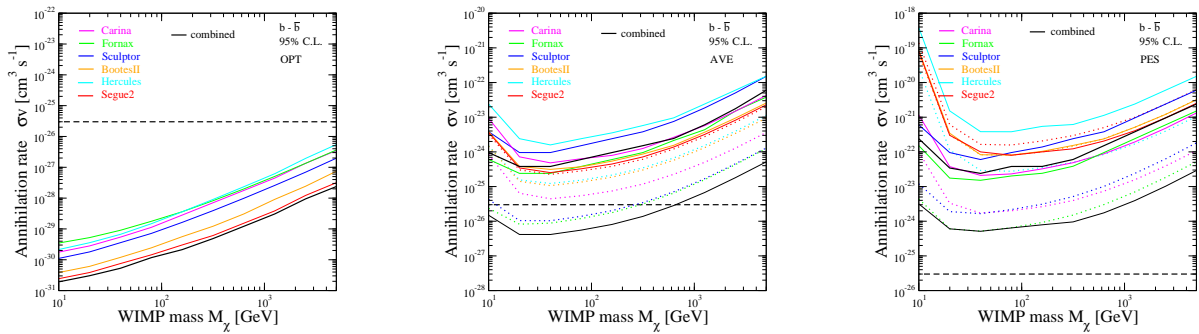

 \begin{minipage}[htb]{5.5cm}
   \centering
   \includegraphics[width=0.8\textwidth]{mvssigmav_bb_OPT.eps}
 \end{minipage}
\hspace{0mm}
 \begin{minipage}[htb]{5.5cm}
   \centering
   \includegraphics[width=0.8\textwidth]{mvssigmav_bb_AVE.eps}
 \end{minipage}
\hspace{0mm}
 \begin{minipage}[htb]{5.5cm}
   \centering
   \includegraphics[width=0.8\textwidth]{mvssigmav_bb_PES.eps}
 \end{minipage}
    \caption{95\% C.L. upper bounds on the velocity averaged annihilation cross-section as a function of the WIMP mass for the tree benchmark scenarios summarized in Table~\ref{tableB}. The annihilation final state is taken to be $b- \bar b$. The bounds in the left panel have been obtained taking the most constraining map between the tapered and un-tapered maps. In the central and right panels, the bounds come from the tapered map with source subtracted in the visibility plane (for the dotted lines, we further subtract sources in the image plane). }
\label{fig:ann1}
 \end{figure} 

As clear from Fig.~\ref{fig:DMsize}, the scenario with loss at injection place can be probed by both the tapered and high-resolution maps, if the profile has an NFW or Einasto shape.
For bounds on the OPT scenario, we will use the map which provides the strongest constraint (since, in this case, low and high resolution maps provide comparable bounds), and we do not subtract point-sources.
For the other two benchmark scenarios, AVE and PES, the emission is on scales $\gtrsim\,1'$, so the bounds can be only obtained from the tapered maps. In these cases, as already mentioned, we can consider two types of approaches and maps.
The more conservative approach relies on maps (MAP1) where the subtraction of sources is performed in the visibility plane only and involves sources from the un-tapered maps (we remind that the un-tapered maps have a synthesized beam of about 6 arcsec and largest detectable scale of 30 arcsec). 
In the more challenging approach, we consider MAP2 which are built from MAP1 by further subtracting point-like sources in its image plane. In the tapered maps, the synthesized beam is $\sim1'$, so here for point-like structures we refer to sizes $\lesssim\,1'$.

\begin{figure}[t]
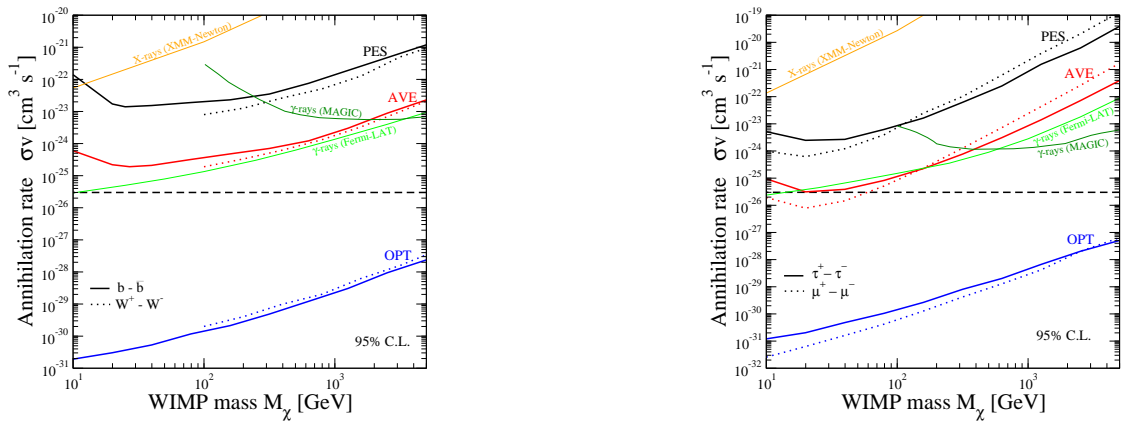

 \begin{minipage}[htb]{7.cm}
   \centering
   \includegraphics[width=0.8\textwidth]{mvssigmav_quark.eps}
 \end{minipage}
\hspace{20mm}
 \begin{minipage}[htb]{7cm}
   \centering
   \includegraphics[width=0.8\textwidth]{mvssigmav_lept.eps}
 \end{minipage}
    \caption{95\% C.L. upper bounds on the velocity averaged annihilation cross-section as a function of the WIMP mass, for the three scenarios  OPT (blue), AVE (red), and PES (black) listed in Table~\ref{tableB}. In the AVE and PES cases, bounds are derived from the tapered maps with sources conservatively subtracted in the visibility plane only.
Left: Annihilations into $b-\bar b$ (solid) and $W^+-W^-$ (dotted). We show also the bounds from the analysis of dSphs in Refs.~\cite{Ackermann:2013yva} (light-green) and \cite{Aleksic:2013xea} (dark-green) at gamma-ray frequencies, and in Ref.~\cite{Jeltema:2008ax} (orange, AVE scenario) in the X-ray band, for the $b-\bar b$ finale state. Right: Annihilations into $\tau^+-\tau^-$ (solid) and $\mu^+-\mu^-$ (dotted). We show also the bounds from the analysis of dSphs in Refs.~\cite{Ackermann:2013yva} (light-green) and \cite{Aleksic:2013xea} (dark-green) at gamma-ray frequencies, and in Ref.~\cite{Jeltema:2008ax} (orange, AVE scenario) in the X-ray band, for the $\tau^+-\tau^-$ finale state.}
\label{fig:ann2}
 \end{figure}

Results are shown in Figs.~\ref{fig:ann1}, \ref{fig:ann2}, and \ref{fig:dec}a.

In Fig.~\ref{fig:ann1}, we consider annihilation into $b-\bar b$ and derive 95\% C.L. bounds in the three benchmark astrophysical models.
Left panel shows that the OPT scenario is indeed very optimistic.
The large value of the equipartition magnetic field and the cuspy $e^+-e^-$ distribution (resulting from the cuspy DM distribution and the assumed radiative loss at injection place) make the constraints extremely strong. The ``thermal'' annihilation rate $\langle \sigma_a v\rangle= 3\cdot 10^{-26} {\rm cm}^3/{\rm s}$ is confidently excluded by all the six dSph bounds. The caveat here is, in addition to the optimistic assumptions, that the DM center could be offset with respect to the optical center, and one of the detected sources in Table~\ref{tab:sources} can be actually induced by WIMP annihilations, as discussed in Section~\ref{sec:PS}.
The behaviour of the curves (bounds increasing with mass) can be easily understood from Eq.~\ref{eqQ}, and is due to the fact that for a given energy density the number density decreases as the mass increases.
For the specific case of Fig.~\ref{fig:ann1}a, Fornax and Segue2 bounds come from the high resolution maps, while the other bounds arise from the low resolution ones.

As expected, the combined analysis slightly improves the bound obtained in the most constraining case. The closest dSphs (Segue2 and BootesII) provide the strongest bounds.

The central panel of Fig.~\ref{fig:ann1} focuses on the AVE scenario.
Note the upturn at low masses occurring in all the dSphs. This is due to the fact that low mass WIMPs induce $e^+-e^-$ whose distribution might peak at energies below the value corresponding to the synchrotron power peak, which, at 2 GHz, is given by $E\simeq 21\,{\rm GeV}/\sqrt{B_{\mu G}}$ (where $B_{\mu G}$ is the magnetic field in $\mu G$).
The significantly larger magnetic field in the OPT scenario was preventing this upturn.

In the AVE case, the bounds from the CDS becomes at the same level as for the Segue2 and BootesII dSphs. This is due to two reasons. First, the formers have a significantly larger size, and in turn a larger confinement time, which makes the depletion of $e^+-e^-$ due to spatial diffusion less effective. Second, CDS have experienced a more significant star formation, thus have a larger magnetic field $B_{\overline{SFR}}$ (see Table 2 in Paper II).

As already noticed in Paper II, if the source subtraction in the visibility plane was not completely successful (as especially for the Fornax and Sculptor FoVs), the bounds that can be derived from MAP1 and MAP2 can significantly differ, up to two orders of magnitude. This can be noticed comparing the dashed and solid lines in Fig.~\ref{fig:ann1}b.
While in MAP2 the rms noise is only a factor of few lower than in MAP1, the difference in the derived bounds is much larger for dSphs with significant detection of extended emission in MAP1 (due to residuals from point-source subtraction). Indeed, in this case, the constrained flux has to obviously be above the best-fit flux, which is significantly above the rms. On the contrary, in MAP2 no significant deviation from the no-signal case has been found, which means that the best-fit flux for the diffuse emission is close to zero.

For similar reasons, the combined analysis does not improve the constraints in the case of MAP1. Indeed, the likelihood of the cases with large residuals (in particular, of the Sculptor dSph where the detection is at 11-$\sigma$, see Table 3 in Paper II) significantly reduces if a negligible flux is considered, and this drives down the global likelihood for low values of $\langle \sigma_a v\rangle$.
This discussion highlights the importance of subtraction of sources. The confusion limit is indeed one of the greatest obstacles to be overcome in this kind of studies. In this respect, future radio telescopes with high sensitivity and spatial resolution (like, e.g., the SKA and its precursors) will be crucial for the study of the DM nature (see discussion in \cite{Colafrancescoetal_SKA2014}).

In the right panel of Fig.~\ref{fig:ann1}, we show the constraints in the PES scenario.
The discussion is similar to the one reported for the AVE case.
The effect of spatial diffusion is even more pronounced and the larger CDS become more promising than the smaller UDS, even though the latter are closer (see also Fig. 11 in Paper II). The Fornax dSph is the most constraining case.
Note that this trend is the opposite with respect to what found in the OPT scenario (and typically also in gamma-ray searches).

In Fig.~\ref{fig:ann2}, we show bounds on the WIMP annihilation cross section for few different final states of annihilation, $b-\bar b$ and $W^+-W^-$ in the left panel, and $\tau^+-\tau^-$ and $\mu^+-\mu^-$ in the right panel. In order to be conservative, we choose to show bounds making use of MAP1. Assuming the residuals in MAP1 are not due to DM annihilations, we can consider all the annihilation rates which are incompatible with at least one target to be excluded.
The case of $W^+-W^-$ is very similar to the $b-\bar b$ case discussed in Fig.~\ref{fig:ann1} (and reported in Fig.~\ref{fig:ann2} for completeness). Indeed, the two $e^+-e^-$ yields are very similar for electron/positron energy below $0.3\,M_{\chi}$ (with $M_{\chi}>M_W\simeq 80$ GeV to have a kinematic allowed production of $W^+-W^-$), see e.g. Fig. 4 in Ref.~\cite{Regis:2008ij}, namely in the relevant energy range for the synchrotron production.

Annihilations in $\tau^+-\tau^-$ and $\mu^+-\mu^-$ induce instead harder spectra of $e^+-e^-$. Bounds for the leptonic channels are thus more stringent than in the hadronic case at low WIMP masses, while the opposite picture occurs in the TeV range.

Fig.~\ref{fig:dec}a shows 95\% C.L. constraints in the decaying-DM parameter space for the $b-\bar b$ and $\mu^+-\mu^-$ final states. 
Again the softer $e^+-e^-$ spectrum of the hadronic case makes the latter more constraining than the leptonic channel for high WIMP masses and viceversa at low mass.
The bounds in the AVE and PES scenarios differ by a factor which is of the same order of the findings in the annihilating case.
The curve of the OPT scenario is instead much closer to the other two benchmark models, and, overall, the uncertainty is significantly reduced.
The dependence on the square of the DM density profile in the annihilating case (with respect to a linear dependence for decaying DM) amplifies the effect of a cuspy initial distribution, if this is not flattened by spatial diffusion as in the AVE and PES scenarios. The huge overdensity of the injected $e^+-e^-$ makes also the equipartition magnetic field large at the center, and these two points explain why the large discrepancy between OPT and AVE found for annihilating DM is not present in Fig.~\ref{fig:dec}a.

In the case of decaying DM, there is no natural scale to aim as for the ``thermal'' annihilation rate in the annihilating DM framework.
However, much interest was recently devoted to decaying DM models in connection to the possibility of explaining the measured raise in the local cosmic-ray positron fraction at high energy. The green region in Fig.~\ref{fig:dec}a shows the best-fit to the PAMELA excess in the case of $\mu^+-\mu^-$ final state (see, e.g., Ref~\cite{Ibarra:2009dr}), which is the only viable interpretation among the final states considered here.

\section{Comparison with previous works}
\label{sec:compar}
Very few studies have so far attempted at deriving particle DM bounds from radio data in the dSph fields (as proposed in \cite{Colafrancesco:2006he}). 
The analysis most directly comparable to ours has been recently presented in Refs.~\cite{Spekkens:2013ik,Natarajan:2013dsa}.
It is based on radio observations of the Ursa Major II and Willman 1 FoVs at 1.4 GHz with the GBT telescope.
The reported sensitivity (7 mJy/beam after subtracting discrete sources) is about two orders of magnitude weaker than the rms noise level of the ATCA observations considered in this work (at 2 GHz).
For peaked spatial distributions, our bounds are therefore about two orders of magnitude more constraining.
For shallower emission profiles, the bound scaling is, on the other hand, lower than the ratio of the sensitivities since the beam of the GBT observations (10 arcmin) is much larger than the ATCA beam (1 arcmin), and so the former telescope can integrate more signal.
In the cases with significant diffusion considered in Refs.~\cite{Spekkens:2013ik,Natarajan:2013dsa}, the current work improves WIMP constraints by about one order of magnitude.
The actual ratio might be actually slightly larger since in our modeling the magnetic field (diffusion coefficient) decreases (increases) exponentially outside the stellar region while in \cite{Spekkens:2013ik,Natarajan:2013dsa} they are taken to be constant.

For a given diffusion scheme, we can directly compare our radio constraints with X-ray bounds arising from IC scattering on the CMB in dSphs. The only assumption which is present in the radio estimate while being not crucial in the X-ray case concerns the magnetic field strength.
Focusing on the AVE scenario, we found that the bounds at $M_\chi=100$ GeV derived in this work are about four orders of magnitude stronger than those derived in Ref.~\cite{Jeltema:2008ax} (which stem from X-ray observations of the Fornax dSph). They are shown in Fig.~\ref{fig:ann2} in orange for $b-\bar b$ and $\tau^+-\tau^-$ final states, after rescaling  the constraints presented in their Fig. 4 for the diffusion model of the AVE scenario (for simplicity, we employed a linear scaling, correcting from a diffusion coefficient of $1.1\times 10^{27}$ cm$^2$/s to $3.0\times 10^{28}$ cm$^2$/s, see also Fig. 3b in Ref.~\cite{Jeltema:2008ax}).
This means that, with current data, radio observations have a significantly better constraining power than X-rays provided a value of the magnetic field strength $B\gtrsim 0.01\,\mu$G. 

Finally, we can compare results of Fig.~\ref{fig:ann1} with bounds obtained at gamma-ray frequencies. 
The radio signal is, however, much more uncertain, due to the unknown description of spatial diffusion and magnetic field, while neither ingredient is needed in the computation of the prompt gamma-ray signal. The current, most constraining DM bounds from the analysis of dSphs are obtained by the Fermi-LAT~\cite{Ackermann:2013yva} (below about 1 TeV) and MAGIC~\cite{Aleksic:2013xea} (in the TeV regime) collaborations. 
For definiteness, we focus again on the comparison with the AVE scenario. The Fermi-LAT bounds are slightly more constraining than Fig.~\ref{fig:ann1} for $b-\bar b$, while of the same order of magnitude for the $\tau^+-\tau^-$ channel. In the latter case, the MAGIC observations significantly improve the constraints in the region above few hundreds of GeV.
Notice that the IC emission is not included in the work of both \cite{Ackermann:2013yva} and \cite{Aleksic:2013xea}, while (within the AVE scenario) is expected to slightly improve constraints in the TeV regime for the $\tau^+-\tau^-$ channel.

\section{Detection prospects}
\label{sec:prospects}
Radio astronomy is entering a golden era. In the next few years, the Square Kilometre Array (SKA) project will be completed in its Phase-1 building the world's largest radio telescope and the Phase-2 will further improve its sensitivity and spectral range.
Its pathfinders, ASKAP and MeerKAT, are in the final constructing phase. The first continuum images have been already produced by a six antenna test prototype of ASKAP, the Boolardy Engineering Test Array (which will be part of the full ASKAP array), and by a seven antenna precursor of MeerKAT, the KAT-7.
The European new-generation radio interferometer, the LOw-Frequency ARray (LOFAR), has recently started operations and will offer an unprecedented coverage of the low-frequency range from 10 to 240 MHz.
Since the end of 2012, the Karl G. Jansky Very Large Array (JVLA) project has been completed significantly improving the sensitivity of current radio telescopes.

All these instruments can be fruitfully employed for the DM search strategy discussed in this work.
In order to provide quantitative predictions, we focus on few observational setups which will be available in the years to come.

First, one can realize a survey similar to the one presented here, but with the interferometric JVLA array in the Northern sky (so also probing different targets).
The rms sensitivity for an integration time of 1 hour per pointing is about $10\,\mu$Jy in the 1-2 GHz band, and the D-configuration is the best-suited one for the detection of diffuse emissions. The D-configuration consists of 27 antennae with dish diameter of 25 meters arranged along the three arms of a Y-shape and has a maximum and minimum baseline of 1 km and 35 m, respectively.
On the other hand, observations in other configurations with higher resolution (longer baselines) are also needed, in order to provide an accurate mapping of point-sources for the subsequent subtraction; indeed the confusion limit of D-configuration is around $90\,\mu$Jy. An overall gain of at least a factor of 2 in the sensitivity can be confidently expected.

One of the key science projects of ASKAP is the Evolutionary Map of the Universe (EMU).
EMU is a survey project which is about to commence operations and will perform a deep continuum survey at GHz-frequency covering the entire Southern Sky with rms sensitivity of $10\,\mu$Jy. 
ASKAP will have 36 independent beams giving a total field of view of 30 square degrees, and with a resolution of about 10 arcsec.
With the EMU data at hand, it will be possible not only to improve the sensitivity with respect to the project discussed here, but also to dramatically enlarge the number of observed dSphs. Indeed, 14 known dSph Milky-Way satellites will be within the covered area, and we expect this number to be further increased by forthcoming optical surveys, in particular of the Southern sky (e.g., Dark Energy Survey~\cite{Abbott:2005bi}, SkyMapper~\cite{SkyMapper}, Large Synoptic Survey Telescope~\cite{Abell:2009aa}). Indeed, similarly, in the past few years, the Sloan Digital Sky Survey data have more than doubled the number of known dSph satellites in the Northern sky.
Putting together the increase in sensitivity and the larger dSph sample, we can foresee a gain in the constraining power of a factor of 5-10.

The South African MeerKAT radio telescope is currently being built, with science operations in the final 64-antenna configuration expected to start in 2017.
Its FoV will be more limited than the ASKAP one, but will have higher survey speed: deep observations of the most promising dSph targets are probably the best strategy to pursue with such a radio telescope.
With few hours of integration time over the region of a single dSph, a $\leq 1\,\mu$Jy rms level can be achieved at GHz-frequency, increasing the sensitivity to diffuse emission in dSph by a factor of $\sim 50$ with respect to the present observations.

We do not discuss in details the prospects for the SKA precursors working at frequencies much lower than in the observations presented here, such as LOFAR and the Murchison Widefield Array (MWA). Indeed, although similar improvements as for the other pathfinders can be foreseen, a simple rescaling of the computed bounds cannot be adopted and a dedicated study would be in order.

Finally, by 2020,the SKA-1 mid telescope array should be deployed and fully operational, increasing the sensitivity of its precursors by two-orders-of-magnitude.
The SKA-1 mid-Band (350 -1050 MHz) will probably be the most promising frequency range for the majority of WIMP models.
The full SKA-2 phase will bring another factor $\sim 10$ increase in sensitivity and an extended frequency range up to at least 25 GHz. Such frequency coverage increase will also contribute to constrain the DM particle mass by looking at the expected spectral cut-off at $\nu_{max} \approx 37 GHz B_{\mu G} (M_{\chi}/100 GeV)^2$ (see, e.g., Ref.~\cite{Colafrancesco:2000zv}).

More in particular, the SKA expected noise can be estimated with:
\be
\sigma_{rms}=\sqrt{\frac{2}{B\,t_{obs}}}\,k_b\,\frac{T_{sys}}{A_{eff}}\;,
\label{eq:SKA}
\ee
where $B$ is the bandwidth, $t_{obs}$ is the observing time, $k_b$ is the Boltzmann's constant, $T_{sys}$ is the system noise temperature, and $A_{eff}$ is the effective area.
Taking $A_{eff}/T_{sys}= 2\cdot 10^4\,{\rm m^2/K }$~\cite{Carilli:2004} and a bandwidth of 300 MHz at GHz frequency, one obtains, from Eq.~\ref{eq:SKA}, $\sigma_{rms}\simeq 30$ nJy for 10 hours of integration time. This is about a $10^3$ factor of gain in sensitivity with respect to the observations presented here. A further improvement by a factor of 2 can be confidently foreseen due to the larger number of accessible dSph satellites.

The SKA will also have the great advantage to be able to determine the dSph magnetic field via Faraday rotation measurements (and possibly also polarization), provided its strength is around the $\mu$G level (which is the expected level based on SFR arguments, see discussion above). This will make the predictions for the expected signal much more robust and obtainable with a single experimental configuration, which can simultaneously measure both the DM-induced synchrotron emission and the magnetic field.

\begin{figure}[t]
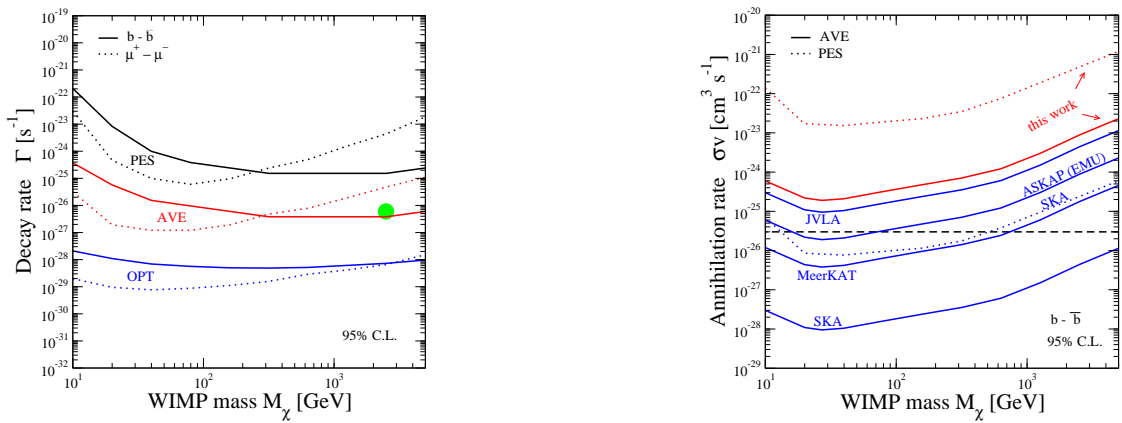

 \begin{minipage}[htb]{7.cm}
   \centering
   \includegraphics[width=0.8\textwidth]{mvsgamma.eps}
 \end{minipage}
\hspace{20mm}
 \begin{minipage}[htb]{7cm}
   \centering
   \includegraphics[width=0.8\textwidth]{mvssigmav_prosp.eps}
 \end{minipage}
    \caption{Left panel: 95\% C.L. upper bounds on the decay rate $\Gamma=1/\tau$ as a function of the WIMP mass, for the three scenarios OPT (blue), AVE (red), and PES (black) listed in Table~\ref{tableB}, and decays into $b-\bar b$ (solid) and $\mu^+-\mu^-$ (dotted). The green circle shows the best-fit to the PAMELA positron fraction for $\mu^+-\mu^-$ final state (see, e.g., Ref~\cite{Ibarra:2009dr}). Right panel: Prospects of detection for WIMPs annihilating into $b-\bar b$ in the AVE and PES scenarios. See text for details about the selected experimental configurations.}
\label{fig:dec}
 \end{figure}

The prospects of detection/constraints of the WIMP particle properties are shown in Fig.~\ref{fig:dec}b for the observing scenarios outlined above, and in the AVE and PES cases.
Note that we can progressively close in on the full parameter space, even in the PES case, up to TeV masses. 

There are two main caveats in the presented forecasts.
The first stems from the fact that, for an extended emission, the confusion issue becomes stronger and stronger as one tries to probe fainter and fainter fluxes.
The source subtraction procedure becomes thus crucial and this can affect the estimated sensitivities. The extent of the impact of this effect on the actual sensitivity is hardly predictable at the present time, especially for the SKA, since it will depend on the properties of the detected sources (in an unknown flux density range), the efficiency of deconvolution algorithms, and the accuracy of the telescope beam shape.

The second caveat is that bringing down the observational threshold, one can start to possibly probe the very low-level of expected non-thermal emission associated to the (very-low) star formation in dSph. A DM contribution should be then disentangled from such astrophysical background.
On the other hand, the superior angular resolution of the SKA will allow to precisely map such two putative emissions, and to correlate them with the stellar or DM profiles (obtained instead via optical and kinematic measurements).

It is clear that a full use of dSph as DM laboratories will require a synergy between optical observations (deep, large-area photometric searches for dSph identification and spectroscopic and astrometric follow-ups to derive structural properties), and multi-frequency observations of non-thermal emissions from radio to gamma-rays.

\section{Conclusions}
\label{sec:concl}
In this paper, we presented a search for a particle DM signal performed observing six dSph galaxies (Carina, Fornax, Sculptor, BootesII, Hercules and Segue2) at radio frequency.
A total of 123 hours of observing time with the ATCA telescope operating at 16 cm wavelength were employed. 
The rms sensitivity of the resulting maps is below 0.05 mJy/beam for all the targets of our observations.

The synchrotron radiation from high-energy electrons and positrons originated from DM annihilations or decays is expected to produce a diffuse signal in dSph satellites of the MW on scales of few arcmin.
We have presented here the first project dedicated to the WIMP search making use of radio interferometers, that could be considered as a pilot experiment for the next generation high-sensitivity and high-resolution radio telescope arrays like the SKA. 

A very recent campaign using instead the GBT single-dish radio telescope was presented in \cite{Spekkens:2013ik,Natarajan:2013dsa}, and similar studies can be conducted with the Effelsberg 100-m single-dish telescope which has the capability of also detecting extremely weak polarised emission.

For the particle DM search we are interested in, the use of multiple array detectors having synthesized beams of arcmin size has a number of advantages with respect to single-dish observations. 
First, the large collecting area can allow to increase the sensitivity.
The best beam choice for the detection of a diffuse emission requires a large synthesized beam (in order to maximize the integrated flux), but still smaller than the source itself to be able to resolve it.
A good angular resolution is also crucial in order to distinguish between a possible non-thermal astrophysical emission and the DM-induced signal, which clearly becomes very hard if the dSph is not well-resolved.
The possibility of simultaneously detecting small scale sources with the long-baselines of the array allows to overcome the confusion limit. In the case of arcmin beams, the confusion level can be easily reached with observations lasting for few tens of minutes, even by current telescopes. A source subtraction is thus a mandatory and crucial step of the analysis.
Finally, single dish telescopes face the additional complication related to Galactic foreground contaminations, which are instead subdominant for the angular scales typically probed by telescope arrays at GHz frequency.
Note that similar arguments apply also for the comparison of capabilities of gamma-ray telescopes versus radio interferometers for what concerns the WIMP search in dSphs. 

No evidence for an extended emission over a size of few arcmin has been detected with current observations (see also discussion in Paper II). We derived bounds on the WIMP annihilation/decay rate as a function of the mass for different final states of annihilation/decay.
They are comparable to the best limits obtained with gamma-ray observations and are much more constraining than what obtained in the X-ray band or with previous
radio observations. 
In Section~\ref{sec:mod}, we described how to model the expected WIMP signal, discussing the involved uncertainties. They are mostly given by the shape of the DM profile and the dSph magnetic properties.
DSphs are poorly known systems and this reflects into a large uncertainty in the predicted signal.
We define an optimistic and a pessimistic scenarios to bracket such uncertainty. The associated bounds on the annihilation rate varies by nearly 7 orders of magnitude in the annihilating DM case (while the variation of the decay rate is more limited in the case of decaying DM). On the other hand, even for the pessimistic scenario, constraints are not dramatically far from the ``thermal'' annihilation rate, and forthcoming observations can be able to cover the full WIMP parameter space.
For an average scenario, where we derive a magnetic field from the SFR of the dSph (averaged over its history), and assume a MW-like spatial diffusion and an NFW profile for the DM distribution, the bounds on the annihilation rate are around the ``thermal'' value for $M_\chi=100$ GeV and leptonic channels, while a factor of 10 above it for hadronic or gauge-boson final states. 
As mentioned, these limits are significantly more constraining the bounds from IC in the X-ray band, and comparable to the Fermi-LAT constraints from DM-induced prompt emission of gamma-rays.

We also investigate the possibility that point-sources detected in the proximity of the dSph optical center might be associated to the emission from a DM cuspy profile.
This possibility is however likely only in the loss at injection scenario while spatial diffusion should in any case flatten the $e^+-e^-$ distribution, making the source extended rather than point-like. 
We found no source spatially coincident with the dSph center.
On the other hand, we have 7 viable candidates with small displacements (around 1 or 2 arcmin) and with no solid identification at other wavelengths. These sources can deserve further investigation since we found that the WIMP scenario can fit the point-like emission with annihilation rates consistent with existing bounds.

To conclude, we demonstrate that radio interferometric observations are a suitable strategy to search for a WIMP-induced diffuse emission in dSphs.
The SKA and its precursor will be able to progressively probe a signal from WIMP scenarios with ``thermal'' annihilation rate and masses up to few TeV, irrespective of  astrophysical assumptions.

\section*{Acknowledgements}
We thank Piero Ullio for insightful discussions during the early stages of the project and Rainer Beck for useful comments. 

S.C. acknowledges support by the South African Research Chairs Initiative of the Department of Science and Technology and National Research Foundation and by the Square Kilometre Array (SKA).
S.P. is partly supported by the US Department of Energy, Contract DE-FG02-04ER41268.
M.R. acknowledges support by the research grant {\sl TAsP (Theoretical Astroparticle Physics)} funded by the Istituto Nazionale di Fisica Nucleare (INFN).

The Australia Telescope Compact Array is part of the Australia Telescope National Facility which is funded by the Commonwealth of Australia for operation as a National Facility managed by CSIRO.

\end{document}